\documentclass[prl,floatfix,twocolumn,showpacs,superscriptaddress]{revtex4}
\usepackage{graphicx}
\def\bea{\begin{eqnarray}}
\def\eea{\end{eqnarray}}
\def\be{\begin{equation}}
\def\ee{\end{equation}}

\def\S{\mbox{\bf S}}

\begin{document}

\author{Didier Poilblanc}
\affiliation{
  Laboratoire de Physique Th\'eorique, CNRS-UMR 5152, Universit\'e Paul Sabatier,
  F-31062 Toulouse, France }

\date{\today}
\title{Enhanced pairing in doped quantum magnets with frustrating hole motion}

\begin{abstract}
Evidence for strong pairing at arbitrary small J/t is provided in
a t-J model on the checkerboard lattice for the sign of hopping
leading to frustration in hole motion. Destructive quantum
interferences suppress Nagaoka ferromagnetism when $J/t\rightarrow
0$ and reduce drastically coherent hole motion in the fluctuating
singlet background. It is shown that, by pairing in various orbital symmetry
channels, holes can benefit from a large gain of kinetic energy.

\end{abstract}
\maketitle

Although the underlying mechanism that leads to high critical
temperature in the cuprates superconductors might be of magnetic
origin a detailed scenario is still missing. The inter-layer
tunnelling mechanism proposed by Anderson~\cite{Interlayer_PWA} is
based on the presence of weakly coupled CuO layers in this
material. The original idea is that the pairing mechanism within a
given layer (which might be rather weak) could be amplified by the
Josephson inter-layer tunnelling of the Cooper pairs. Such a
scenario is only possible if the single particle motion is not
allowed along the c-axis perpendicular to the layers, i.e. if the
coherent single particle tunnelling is blocked by some means. This
phenomenon could be observed e.g. in weakly coupled (along the
c-axis) two-dimensional (2D) Luttinger liquids~\cite{2D-LL} (LL)
as a result of the orthogonality catastrophe~\cite{Chakravarty}.

\begin{figure}
  \centerline{\includegraphics*[width=0.9\linewidth]{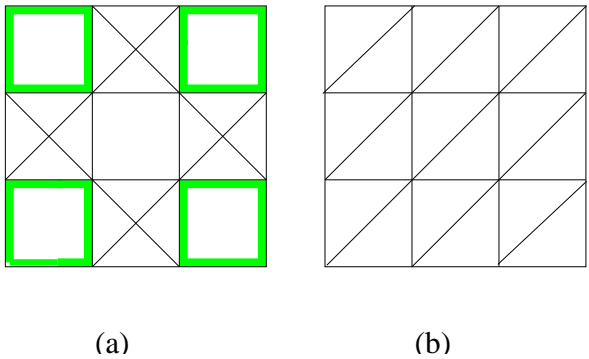}}
  \caption{\label{fig:lattices}
 (a) Checkerboard lattice. The plaquette singlets of the VBS state are shown
 schematically in green. (b) Lattice isomorphic to the triangular lattice
 with the same average connectivity as (a).}
\end{figure}

Despite its beauty and simplicity, this scenario still lacks
strong experimental support and, on the theoretical side, a proof
that an analog of the one-dimensional LL could be realized in 2D.
Nevertheless, I suggest here that a closely related mechanism
could be realized in a doped, strictly 2D, antiferromagnetic (AF)
frustrated magnet. In the last years, there has been growing
experimental and theoretical interests in frustrated magnets which
could provide e.g. realizations of new exotic gapped spin liquid
phases. Transition metal oxides (like A$_2$Ti$_2$O$_7$
titanates~\cite{titanates}) with a pyrochlore type structure, a
three dimensional lattice of corner-sharing tetrahedra, exhibit a
wide range of interesting physical properties. The AF Heisenberg
model defined on the checkerboard lattice~\cite{pyrochlore,Fouet},
a 2D analog of the pyrochlore lattice, is believed to form a
gapped Valence Bond Solid (VBS)~\cite{VBS} with an ordered
arrangement of plaquette singlets on a subset of the void
plaquettes~\cite{Fouet} (see Fig.~\ref{fig:lattices}(a)) and a
rather short magnetic correlation length of only a few lattice
spacings. The GS hence breaks translation symmetry and is two-fold
degenerate.

\begin{figure}
  \centerline{\includegraphics*[width=0.9\linewidth]{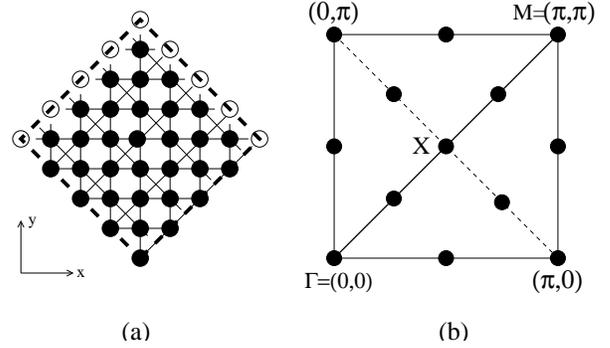}}
  \caption{\label{fig:cluster}
(a) Periodic $\sqrt{32}\times\sqrt{32}$ cluster used in this
study. Note the full $C_{4v}$ point-group symmetry around the
centers of the crossed and void plaquettes. (b) Reciprocal space
and available $\bf k$-points. Note that the checkerboard pattern
of the crossed plaquettes leads to a folding w.r.t the dashed
line.}
\end{figure}

The recent discovery of superconductivity in 5d transition metal
pyrochlores~\cite{5d-pyrochlore} or in a CoO triangular layer
based compound~\cite{cobaltites} suggests that geometric
frustration, which could be magnetic and/or kinetic, might play a
key role in the mechanism of unconventional superconductivity. In
this Letter, I show that pairing appears at arbitrary small $J/t$
in the hole-doped checkerboard AF magnet for a given sign of the
hopping amplitude~\cite{note1} leading to frustration in single
hole motion. It is shown that hole delocalization i.e. {\it gain
in kinetic energy} can play a key role in some mechanisms of
unconventional pairing, as in the inter-layer mechanism discussed
above.

\begin{figure}
  \centerline{\includegraphics*[width=0.96\linewidth]{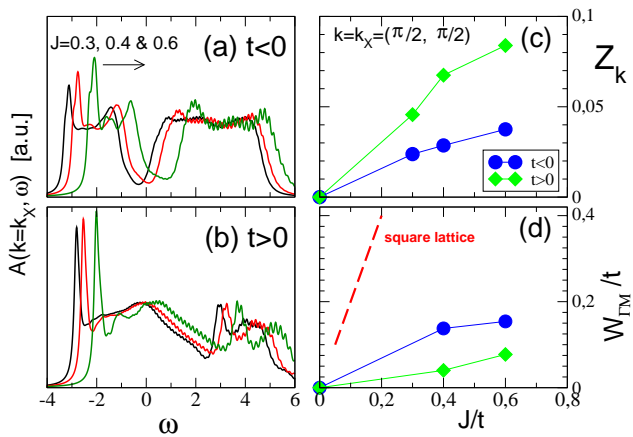}}
  \caption{\label{fig:spectralX_vsJ}
Single hole dynamics on the checkerboard lattice;  Spectral
functions for $t<0$ (a) and $t>0$ (b). The total weight is a
constant (sum-rule) and set to 1; (c) Z-factor at the X-point vs
$J/t$; (d) Dispersion width along the $k_x=k_y$ direction in the
BZ. For comparaison the case of the square lattice is shown.}
\end{figure}

To describe the low-energy physics of the weakly doped magnet the
standard $t{-}J$ model Hamiltonian is used:
  \begin{equation}
   H= - t \sum_{i,j;\sigma}\ \mathcal{P}
      c^{\dagger}_{i,\sigma}c_{j,\sigma}\mathcal{P}
    + J \sum_{\langle i,j\rangle}\ \S_i \cdot \S_j -\frac{1}{4} n_i n_j
  \end{equation}
where the same couplings $t$ and $J$ are used on the vertical,
horizontal and diagonal bonds of the checkerboard lattice.
Computations are performed by Lanczos exact diagonalisation of a
32-site periodic cluster. Although tilted by 45$^o$ w.r.t. the
crystallographic axis, the square $\sqrt{32}\times\sqrt{32}$
cluster (see Fig.~\ref{fig:cluster}(a)) exhibits the full $C_{4v}$
point group symmetries of the infinite lattice (with 2 sites per
unit cell) so that all orbital symmetries (s, $d_{x^2-y^2}$,
$d_{xy}$, etc...) can be considered~\cite{note2}. In addition, the
most symmetric $\bf k$-points are available in the Brillouin zone
(BZ) as shown in Fig.~\ref{fig:cluster}(b). Note that, since the
unit cell contains 2 sites, a folding of the Brillouin zone occurs
so that the many-body spectra at $\bf k$ and $\bf k+Q_0$ ($\bf
Q_0$ is the AF wavevector) are identical (although spectral
weights are different). An additional folding is expected {\it in
the thermodynamic limit} due to the translation symmetry breaking
of the VBS GS.

\begin{figure}
  \centerline{\includegraphics*[width=0.9\linewidth]{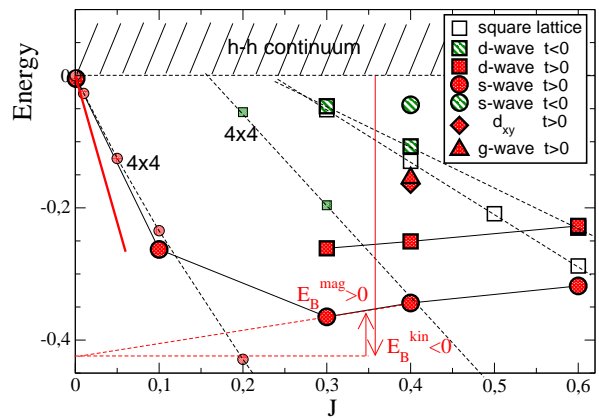}}
  \caption{\label{fig:ener_2h}
   Energies of two holes doped on a square 32-site cluster
($\sqrt{32}\times\sqrt{32}$) in various symmetry channels
(as indicated).
 The energy reference ($E_B=0$) is chosen as the bottom of the
two-hole continuum. Boundstates are characterized by $E_B<0$. The
thick red line corresponds the local (negative) slope at
$J/t\rightarrow 0$. Thin red arrows show the large (small) kinetic
(magnetic) energy gain (lost) obtained from the estimated slope of
$E_B$ vs $J$ at $J\simeq 0.35$ (see text). Data obtained on a
$4\times 4$ cluster are also shown using smaller symbols.}
\end{figure}

Let me first briefly review the properties of a single hole doped
into the VBS GS of the checkerboard lattice (see
Fig.~\ref{fig:spectralX_vsJ}). Numerical computations based on the
analysis of the hole Green function~\cite{Laeuchli-Poilblanc} (see
also Figs.~\ref{fig:spectralX_vsJ}(a-b)) show the existence of
quasiparticle (QP) poles~\cite{note3} whose Z-factor increases
with $J/|t|$ as shown in Fig.~\ref{fig:spectralX_vsJ}(c). The mass
of the single hole can be estimated e.g. from the dispersion of
its QP pole along the $k_x=k_y$ direction of the Brillouin zone
(for which the 32-site cluster contains 5 points). As seen in
Ref~\cite{Laeuchli-Poilblanc} and shown in
Fig.~\ref{fig:spectralX_vsJ}(d) the single hole bandwidth is much
smaller than in the (gapless) square lattice where $W\sim
2.2J$~\cite{Laughlin}. In the latter case, the strong
renormalization of the coherent motion can be explained e.g. by
long-wavelength spin-waves scattering~\cite{Laughlin}. In
contrast, in the checkerboard lattice with $t>0$ the QP is
extremely massive, with a very narrow bandwidth, although the VBS
host has no low energy excitations. This results from a subtle
interplay between the intrinsic frustrated nature of the hopping
and the long-ranged plaquette order induced by magnetic
frustration. Indeed, one expects destructive
interferences~\cite{localisation} between the paths available for
the hole to hop {\it coherently} from one plaquette to the next.
However, {\it incoherent} motion still occurs. In that respect,
the behavior when $J/t\rightarrow 0$ (for which exact statements
can be given) is very instructive: while for $t<0$ the GS is a
Nagaoka ferromagnet (of energy $-6|t|$), for $t>0$ the $S=1/2$ GS
remains stable down to $J=0$ where its energy becomes $-4t$ (much
lower than the fully polarized state of energy $-2t$ where the
hole is localized on a plaquette). Interestingly enough, the
triangular lattice of Fig.~\ref{fig:lattices}(b) exhibits the very
same behavior at $J/t\ll 1$.

As shown below, I argue that holes doped into the checkerboard AF
magnet pair up very strongly when $t>0$. The pairing is studied
here by considering the same 32-site cluster doped with two holes.
The two hole GS energies are computed for all possible orbital
symmetries. The hole-hole binding energies $E_B$ obtained by
subtracting twice the single hole GS energy are plotted in
Fig.~\ref{fig:ener_2h}. The results for $t>0$ and $t<0$ reveal
striking differences which point toward two completely different
physical behaviors. The case $t<0$ shows a behavior qualitatively
(even quantitatively) similar to the case of the well-known square
lattice~\cite{hh_square} (for which the sign of $t$ is
irrelevant): (i) the orbital symmetry of the BS is
$d_{x^2-y^2}$-wave and (ii) the binding energy scales like $J$
(above a size dependent critical value) as expected for a purely
magnetic mechanism. In contrast, the $t>0$ case shows a completely
new behavior: (i) the binding energy appears {\it immediately} at
infinitesimal $J$; (ii) finite size effects in this regime are
well controlled: binding is vanishing at $J=0$ (for all sizes) and
the magnitude of the (negative) slope characterizing the linear
behavior $E_B\propto J/t$ {\it increases} with system size; (iii)
the lowest BS has s-wave symmetry; (iv) other exotic stable BS
appear with non-conventional orbital symmetries like
$d_{x^2-y^2}$-, $d_{xy}$- or even $g$-wave symmetries. It should
be noted that, although the s-wave BS wavefunction has the full
symmetry of the crystal by definition, it could be very
anisotropic like the superconducting gap in the inter-layer
scenario~\cite{Interlayer_PWA}. Importantly enough, while the BS
stability in the thermodynamic limit at small $J/t$ values (the
physical range of this parameter) remains controversial in the
case $t<0$ (as for the square lattice~\cite{hh_square}), the data
(and their behavior with size) for $t>0$ provide a very strong
evidence for the robustness of the BS down to $J=0$. Such a
remarkable phenomenon is due to the nature of the pairing
enhancement mechanism which always benefit from a gain in kinetic
energy. Indeed, using Feynman-Hellmann theorem, the change of
magnetic energy in the BS is given by $E_B^{\rm
mag}=J\frac{dE_B}{dJ}$ and hence can be directly estimated from
the local slope of the function $E_B(J)$ using the simple
construction shown in Fig.~\protect\ref{fig:ener_2h}. Hence, the
difference $E_B^{\rm kin}=E_B-E_B^{\rm mag}<0$ ($\propto -J^2/t$
when $J/t\rightarrow 0$) should be of kinetic origin. It is
remarkable that, at small intermediate $J/t$ couplings (typically
$\sim 0.2-0.6$ for 32 sites), the stability of the BS benefits
only from a large gain of kinetic energy while an overall magnetic
energy is lost in the pairing process. In that regime, the
frustrating AF magnetic coupling is essential to stabilize the
plaquette VBS structure. Therefore, one expects that such a
behavior appears for even smaller $J/t$ as singlet plaquette
correlations build up for larger clusters. Then, one cannot
completely exclude the fact that the $J=0$ limit becomes singular.

\begin{figure}
 \centerline{\includegraphics*[width=0.9\linewidth]{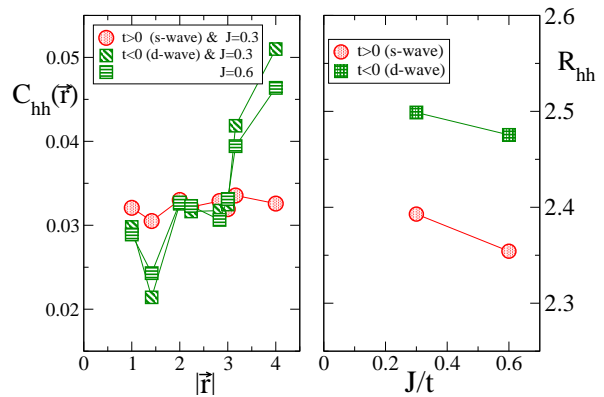}}
\caption{\label{fig:hh_cor} Hole-hole density static correlation
function
 $C_{hh}({\bf r})$ vs $|{\bf r}|$ in the 2-hole GS for both signs of
$t$ computed on the 32-site periodic cluster; (a) Correlations vs
distances for $J/t=0.3$ \& $J/t=0.6$. For $t>0$, the data at
$J=0.6$, close to the one shown here for $J/t=0.3$, are omitted
for clarity; (b) Average hole-hole distance $R_{hh}$ vs $J/|t|$
for both signs of $t$.}
\end{figure}

To investigate further the structure of the two hole BS
wavefunctions it is of interest to compute the static hole-hole
correlation $C_{hh}({\bf r})$, i.e. the probability to find the
two holes at a given separation $\bf r$. Results are shown in
Fig.~\ref{fig:hh_cor}(a) as a function of distance. The data for
$t<0$ show a short-distance "depression" in the hole-hole
correlations and a higher probability for the holes to sit at
distances $\simeq 3$ or $4$. As expected, the size of the
pair wavefunction
tends to grow as $J/|t|$ is reduced (similarly to the case of the
square lattice~\cite{hh_square}) as seen also from the behavior of
the average hole-hole separation $R_{hh}=\sqrt{\big<{\bf
r}^2\big>}$ plotted w.r.t $J/|t|$ in Fig.~\ref{fig:hh_cor}(b).
Below a critical $J/|t|$ value, the d-wave BS might then be
unstable. In contrast, for $t>0$, the hole-hole correlations in
the s-wave BS are rather constant with distance showing no
tendency for the two holes to repel each other (even for small
$J/t$ values). The fact that a strong signature of an effective "attraction"
in the quantities plotted in Figs.~\ref{fig:hh_cor}(a) \& (b)
is missing is probably directly connected to the origin of the pairing
which is of kinetic nature and, hence, should naturally lead to larger pair
sizes. Incidently, one can also expect that such pairing states
are likely to be less sensitive to extended Coulomb repulsion.

From the numerical data presented above a simple scenario might be
drawn for $t>0$. Although its coherent motion is very much
suppressed, a single hole bears a large incoherent motion (see
e.g. the large low-energy incoherent weight in
Figs.~\ref{fig:spectralX_vsJ}(a-b)) and, hence, can melt the
plaquette VBS in its vicinity. This region, which might be fairly
extended in space, becomes more favorable for a second hole to
gain kinetic energy leading to correlated (or assisted) hopping.
It is interesting to point out similarities with the case of some
frustrated tight binding lattices showing both (i) single particle
states localised in so-called Aharonov-Bohm cages~\cite{AB_cage}
and (ii) interaction-induced delocalized two-particle
BS\cite{correlated_hopping}.

\begin{figure}
 \centerline{\includegraphics*[width=0.7\linewidth]{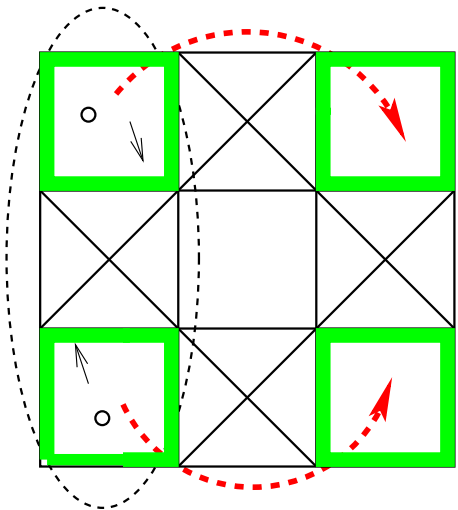}}
  \caption{\label{fig:coherence_new}
Schematic correlated pair hopping process in the checkerboard AF
magnet as indicated by red arrows. The plaquette units of the VBS
are shown in green. Note that the pair wavefunction extends on
several units (see text).}
\end{figure}

In this Letter, I give evidence for two very different behaviors
regarding hole pairing in the t-J model depending whether hole
motion is frustrated or not, i.e. depending on the sign of the
hole hopping on a frustrated lattice such as the checkerboard
lattice. For $t<0$ and sufficiently large $J/|t|$ values pairing
originates from a local magnetic effective attraction analogous to
the case of the non-frustrated square lattice. On the contrary,
for $t>0$, pairing occurs at arbitrary small $J/t$. The stability
of the pair BS always benefits from a large gain of kinetic energy
which might even overcome lost of magnetic energy in some regime.
In this case, the binding energy reaches a few tenths of $t$ so
that high superconducting critical temperature would be expected.
Interestingly enough, pairing occurs in several pairing channels
with different orbital symmetries. This mechanism bears
similarities with the inter-layer mechanism proposed for the high
temperature cuprate superconductors~\cite{Interlayer_PWA}. Note
also that, at finite hole density, a paired state with an enhanced
kinetic energy gain is very likely to be lower in energy than a
phase separated state stabilized only by a magnetic energy gain.
Lastly, I notice that triangular and checkerboard
lattices behave in a very similar way for $J/t\ll 1$.

I thank IDRIS (Orsay, France) for allocation of CPU-time on the
NEC-SX5 supercomputer. I am indebted to A.~L\"auchli for help in
checking some preliminary numerical results. I also acknowledge
stimulating discussions with S.~Capponi, M.~Mambrini and
D.J.~Scalapino.


\begin{thebibliography}{}
\bibitem{note2} Finite size effects are estimated by comparing
these data to those obtained on a smaller $4\times 4$ cluster.
\end{thebibliography}
\end{document}